\documentclass[10pt,journal]{IEEEtran}
\usepackage{epsfig,rotating,setspace,latexsym,amsmath,epsf,amssymb,amsfonts,bm,theorem,cite,authblk, bbm,color, hyperref, multirow, caption, subcaption}

\IEEEoverridecommandlockouts
\allowdisplaybreaks

\title{Is Semantic Communications Secure? \\ A Tale of Multi-Domain Adversarial Attacks}
\begin{document}
\author[1]{Yalin E. Sagduyu}
\author[1]{Tugba Erpek}
\author[2]{Sennur Ulukus}
\author[3]{Aylin Yener}

\affil[1]{\normalsize  Virginia Tech, Arlington, VA, USA}

\affil[2]{\normalsize University of Maryland, College Park, MD, USA}

\affil[3]{\normalsize  The Ohio State University, Columbus, OH, USA}
\maketitle
\begin{abstract}
Semantic communications seeks to transfer information from a source while conveying a desired meaning to its destination. We model the transmitter-receiver functionalities as an autoencoder followed by a task classifier that evaluates the meaning of the information conveyed to the receiver. The autoencoder consists of an encoder at the transmitter to jointly model source coding, channel coding, and modulation, and a decoder at the receiver to jointly model demodulation, channel decoding and source decoding. By augmenting the reconstruction loss with a semantic loss, the two deep neural networks (DNNs) of this encoder-decoder pair are interactively trained with the DNN of the semantic task classifier. This approach effectively captures the latent feature space and reliably transfers compressed feature vectors with a small number of channel uses while keeping the semantic loss low. We identify the multi-domain security vulnerabilities of using the DNNs for semantic communications. Based on adversarial machine learning, we introduce test-time (targeted and non-targeted) adversarial attacks on the DNNs by manipulating their inputs at different stages of semantic communications. As a computer vision attack, small perturbations are injected to the images at the input of the transmitter's encoder.  As a wireless attack, small perturbations signals are transmitted to interfere with the input of the receiver's decoder. By launching these stealth attacks individually or more effectively in a combined form as a multi-domain attack, we show that it is possible to change the semantics of the transferred information even when the reconstruction loss remains low. These multi-domain adversarial attacks pose as a serious threat to the semantics of information transfer (with larger impact than conventional jamming) and raise the need of defense methods for the safe adoption of semantic communications.         
\end{abstract}


\section{Introduction} \label{sec:Intro}
Conventional communications pursues the goal of reliable transfer of messages in terms of symbols (or bits) without a special focus on how the semantics of information pertinent to these messages is preserved. \emph{Semantic communications} seeks to change this paradigm by preserving the semantics in recovered messages beyond conventional reliability measures. 
 As an example, consider a surveillance system of edge devices equipped with cameras. Each edge device takes images and needs to transfer them over a wireless channel to a security center. The semantics of the transferred information is of paramount importance. For example, if the security center needs to classify images with respect to the intruders detected, the goal would be not only to reconstruct the images reliably at the security center, but also to preserve their semantics, namely minimize the semantic loss with respect to the errors in image classification over wireless links to detect intruders and generate alerts at the security center.  

Semantic communications aims to reliably communicate the meanings of messages through a channel by minimizing the semantic error \cite{guler2014semantic} to best preserve the meaning of recovered messages. Semantic communications is envisioned to serve different applications such as \emph{text} \cite{guler2018semantic, xie2021deep}, \emph{speech/audio} \cite{weng2021semantic, walidaudio}, \emph{image} \cite{qin2021semantic} and \emph{video} \cite{Geoffreyvideo} communications. To that end, semantic communications has been studied in terms of information-theoretical foundations \cite{gunduz2022beyond}
and networking aspects \cite{uysal2021semantic}. In addition, \emph{task-oriented communications} has been formulated to utilize the semantics of information via its significance relative to the goal of information transfer when performing an underlying task \cite{shao2021learning}.

In conventional communications, the transmitter and receiver functionalities are typically designed as separate communication blocks such as source coding, channel coding, and modulation at the transmitter and demodulation, channel decoding and source decoding at the receiver. The goal of conventional communications is to reconstruct the transmitter's data samples (messages) at the receiver by minimizing the symbol/bit error rate or a signal distortion metric such as the mean squared error (MSE). The joint design of communication functionalities is ultimately needed to recover the semantic information in addition to the transfer of messages themselves. 

For that purpose, semantic communications can be set up in a deep learning-driven end-to-end communication framework by training an \emph{autoencoder} that consists of an \emph{encoder} at the transmitter and a \emph{decoder} at the receiver. The encoder is modeled as a deep neural network (DNN) for joint operations of source coding, channel coding, and modulation at the transmitter. Then, a second DNN is used to model the \emph{decoder} for joint operations of demodulation, channel decoding, and source decoding at the receiver. The input of the encoder is of general data type  (e.g., an image). The encoder and the decoder are separated by a wireless channel such that the output of the encoder is a modulated signal transmitted over a wireless channel and the received signal becomes the input to the decoder. Then, the decoder output (namely, the reconstructed samples) is given as input to a \emph{semantic task classifier} (the third DNN) that aims to verify the semantics of reconstructed samples (e.g., presence or absence of the intruder in the surveillance scenario described above). If the accuracy of this semantic task classifier is high, then we can say that the semantics of the information is preserved with high fidelity.  

The autoenconder is trained by accounting for channel effects as well as preserving the semantics of information. To that end, semantic communications extends \emph{autoencoder communications}, where the encoder encompasses channel coding and modulation operations, the decoder encompasses demodulation and channel decoding operations, and the sole goal is the same as conventional communications, namely reconstructing messages (in form of symbols) at the receiver by considering channel effects \cite{Oshea1}. On the other hand, the autoencoder for semantic communications incorporates source coding and source decoding, and reconstructs the input data samples. More importantly, this autoencoder is trained by a \emph{custom loss function} that augments the \emph{reconstruction loss} (e.g., the MSE between the input image at the transmitter and the reconstructed image at the receiver) with the \emph{semantic loss} that is represented by the penalty of violating the constraint that the loss of the semantic task classifier (that is designed to capture the semantics of the recovered information) exceeds a target threshold.

The training of the autoencoder and the semantic task classifier can be either fully separated, or better combined together in terms of their input/output relationships and loss functions. The former way is to consider a fixed task classifier that is trained offline (such as in \cite{zhang2022deep}). However, if this classifier is trained with clean input data without taking the channel effects and the corresponding reconstruction losses into account, it cannot achieve high accuracy especially in the low signal-to-noise ratio (SNR) regime. On the other hand, the autoencoder's output is not known in advance before training it with respect to channel effects, so it cannot be readily used as the input to the semantic task classifier for offline training purposes.

To that end, we consider \emph{interactive training} of the autoencoder and the semantic task classifier such that they are retrained over multiple rounds. In each round, the autoencoder for semantic communications is trained first and then its output is used to build the training data that is leveraged to train the semantic task classifier. Along with the reconstruction loss (the MSE loss), the loss of this classifier is then used in the custom loss function of the autoencoder for the next round. This process is repeated over multiple rounds while ingesting new training and validation data samples in each round. This training process seeks to improve the fidelity of both the autoencoder and the semantic task classifier for semantic communications. 

As deep learning becomes a core part of semantic communication systems, there is an increasing concern about the vulnerability of the underlying DNNs to adversarial effects. In our case, three DNNs are utilized, an encoder at the transmitter and a decoder and classifier at the receiver. Smart adversaries may leverage emerging machine learning techniques to exploit vulnerabilities and tamper with the learning functionalities of all these DNNs embedded in semantic communications. The problem of learning in the presence of adversaries has been studied under \emph{adversarial machine learning} for various data domains such as computer vision and natural language processing (NLP). Due to the shared and open nature of wireless medium, wireless applications are highly susceptible to adversaries such as eavesdroppers and jammers that can further observe and manipulate the training and test (inference) processes of machine learning used for wireless applications \cite{adesina2022}. 

In test time, an \emph{adversarial (evasion) attack} can add a small perturbation to the input samples of a victim DNN and fool it into making wrong decisions. The complex decision of the DNN makes it highly sensitive to even small variations in the input samples. Both the encoder at the transmitter and the decoder at the receiver take inputs that can be manipulated by the adversaries. The input of the encoder is a signal of general type. If it is an image such as in the surveillance scenario discussed earlier, the adversary can position a small deceptive object in front of the camera and this is captured as a small perturbation when the camera is taking an image. This corresponds to a computer vision attack. On the other hand, the input of the decoder is the wireless signal received over the channel. This signal can be manipulated by the adversary that transmits a perturbation signal over the air. This way, the received signal includes the perturbation signal superimposed with the transmitted signal and receiver noise. Adversarial attacks on wireless signals have been considered for both signal classifications tasks \cite{kim2021channel} and autoencoder communications \cite{sadeghi2019physical}. In both computer vision and wireless attack cases, the adversarial perturbations are determined as solutions to an optimization problem to minimize the power of perturbation signal subject to the condition that the DNN's decision is incorrect. 

The adversary can launch either a \emph{non-targeted attack} (where the adversary seeks to change the semantics of recovered information to any other incorrect meaning) or a \emph{targeted attack} (where the adversary seeks to change the semantics of recovered information to a specific incorrect meaning).  All these attacks are very effective even when a small perturbation is used, and significantly outperform conventional attacks such as jamming attacks, where the adversary transmits a Gaussian signal as the perturbation. Among adversarial attacks, the non-targeted attack is easier in terms of its objective, and becomes more effective than the targeted attack on the average.          

We also present a \emph{multi-domain attack} framework, where these stealth adversarial attacks can be launched either separately or together against semantic communications by adding perturbations to the input data samples (e.g., images) as well as to the channel data (with over-the-air transmissions). We show that the adversarial attacks can be very effective individually or better when combined in a multi-domain attack. In particular, the addition of an adversarial attack from the computer vision on top of a wireless attack (each using only a small perturbation added to the input image or the wireless signal) can substantially decrease the performance of semantic communications beyond what a single-domain attack can individually achieve.  

The rest of the paper is organized as follows. Section \ref{sec:EndtoEndSemantic} describes the deep learning-driven autoencoder-based semantic communications system and evaluates its performance in terms of reconstruction and semantic losses. Section \ref{sec:Attacks} presents the multi-domain adversarial attack vectors against semantic communications and shows that these attacks can cause a major loss in preserving the semantic information beyond the reconstruction loss. Section \ref{sec:Conclusion} concludes the paper. 

\section{Semantic Communications with End-to-end Deep Learning} \label{sec:EndtoEndSemantic}
\begin{figure}[h]
\centering
\includegraphics[width=\columnwidth]{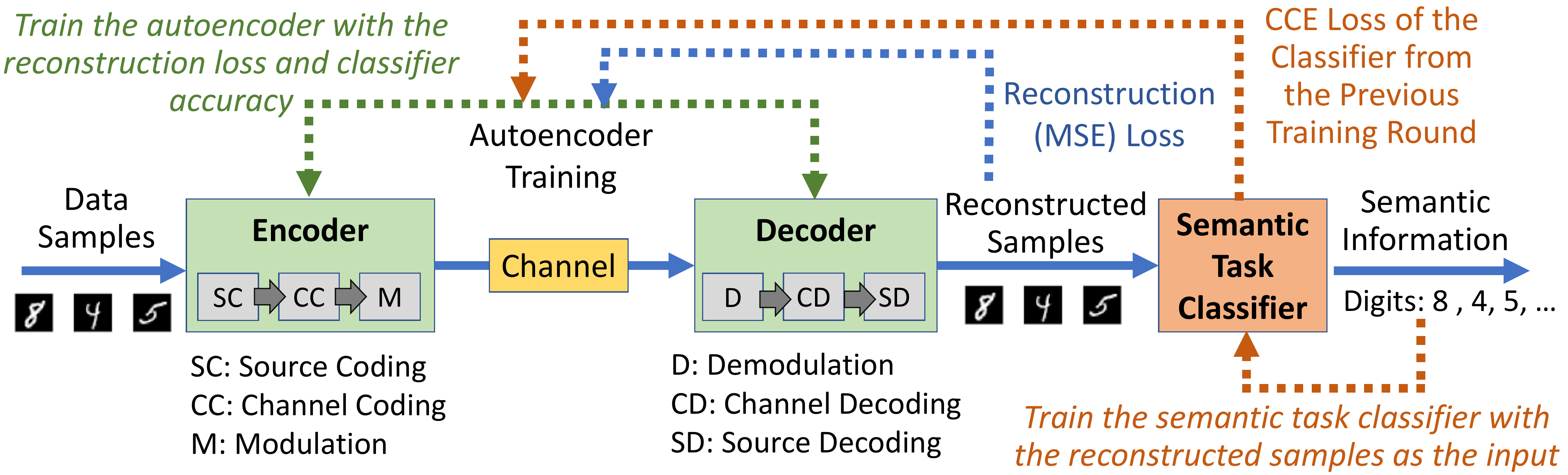}
\caption{System model for semantic communications.}
\label{fig:system}
\end{figure}

We consider a deep learning-enabled semantic communications system shown in Fig. \ref{fig:system}. The transmitter-receiver functionalities are designed as an autoencoder. The encoder trained as a DNN at the transmitter takes the data samples (e.g., images) as the input and jointly performs source coding, channel coding and modulation operations. The output of the encoder is the modulated signals that are transmitted over the air in multiple channel uses. The number of channel uses is equal to the size of latent space at the output of the encoder. The decoder trained as another DNN at the receiver takes the received signals as the input and jointly performs demodulation, channel decoding and source decoding operations. The output of the decoder is the reconstructed data samples.

The encoder and decoder DNNs are trained jointly. Autoencoder communications has been considered for joint training of channel coding and modulation at the transmitter and demodulation and channel decoding at the receiver to minimize the categorical cross-entropy (CCE) loss for symbol recovery \cite{Oshea1}. In our setting, the input data is a general signal such as an image instead of symbols. Therefore, source coding and source decoding are added for the encoder and decoder, respectively. Then, the goal is extended to reconstructing the input signal at the receiver. To construct data samples at the receiver, a distortion loss can be minimized such as the mean squared error (MSE). 

In semantic communications, the goal is not only to reconstruct the signals at the receiver but also preserve the meaning conveyed by the reconstructed data samples by minimizing a semantic loss. We represent this semantic loss with respect to the loss of the semantic task classifier that takes  the reconstructed signals as the input and verifies the semantics of these signals by checking the output of the semantic task classifier. For numerical results, we use the MNIST dataset of handwritten digits as the input data and the digit classifier as the semantic task classifier such that the goal of semantic communications is to ensure that the reconstructed signals can be still reliably recognized with respect to its digit labels.     

To train the autoencoder for semantic communications, we define a custom loss that is the MSE of reconstructed signals augmented with the penalty of violating the condition that the target loss of the semantic task classifier exceeds a certain threshold, namely, the loss can be represented by the MSE loss plus a weight times the gap of classifier loss from a threshold. For numerical results, we set the weight as 0.2 and the threshold as the loss of classifier with clean inputs in the absence of channel effects.

\begin{figure}[h]
\centering
\includegraphics[width=\columnwidth]{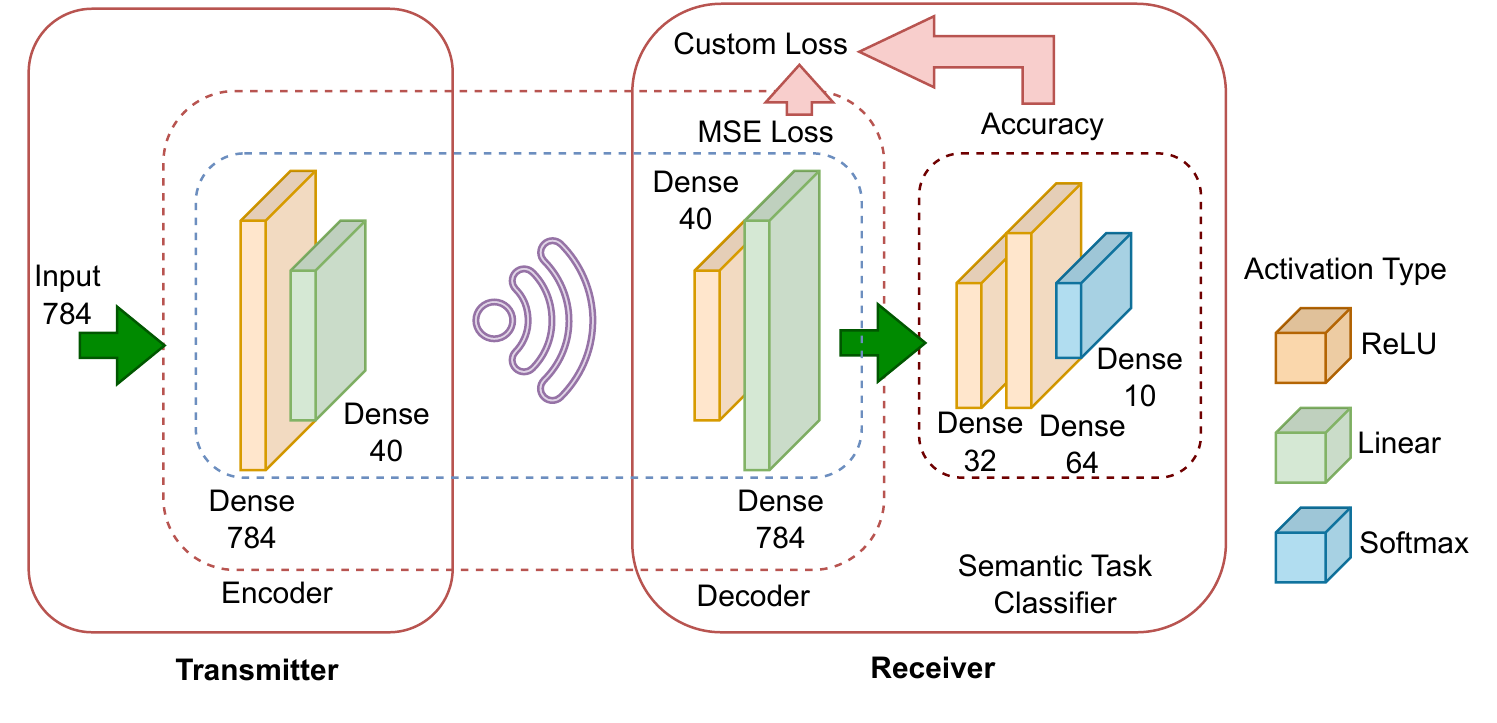}
\caption{The DNN architectures of the autoencoder and the semantic task classifier.}
\label{fig:NNarchs}
\end{figure}

First, we can assume that the DNN classifier that performs the semantic task is pretrained using the CCE loss for the clean data (without channel impairments) as in \cite{zhang2022deep}. However, the semantic task classifier takes the reconstructed signals as the input and is sensitive to the noise in the input data due to channel effects. This data mismatch in training and test times leads to a decrease in the accuracy of the semantic classifier and consequently a decrease in the performance of the autoencoder that makes use of the CCE loss of this classifier as part of its custom loss function as described earlier. To mitigate this issue and improve the training of the autoencoder, we pursue a multi-round interactive training process as follows. In each round, the autoencoder is retrained by using both the reconstruction loss (MSE) and the CCE loss of the semantic task classifier that was trained in the previous round. Then, the semantic task classifier is retrained using the reconstructed samples collected at the output of the decoder in test time of the current round. Starting with pretraining of the semantic task classifier with clean data, we repeat this process in multiple rounds to improve the overall performance of semantic communications.  

\begin{figure}[h]
\centering
\includegraphics[width=\columnwidth]{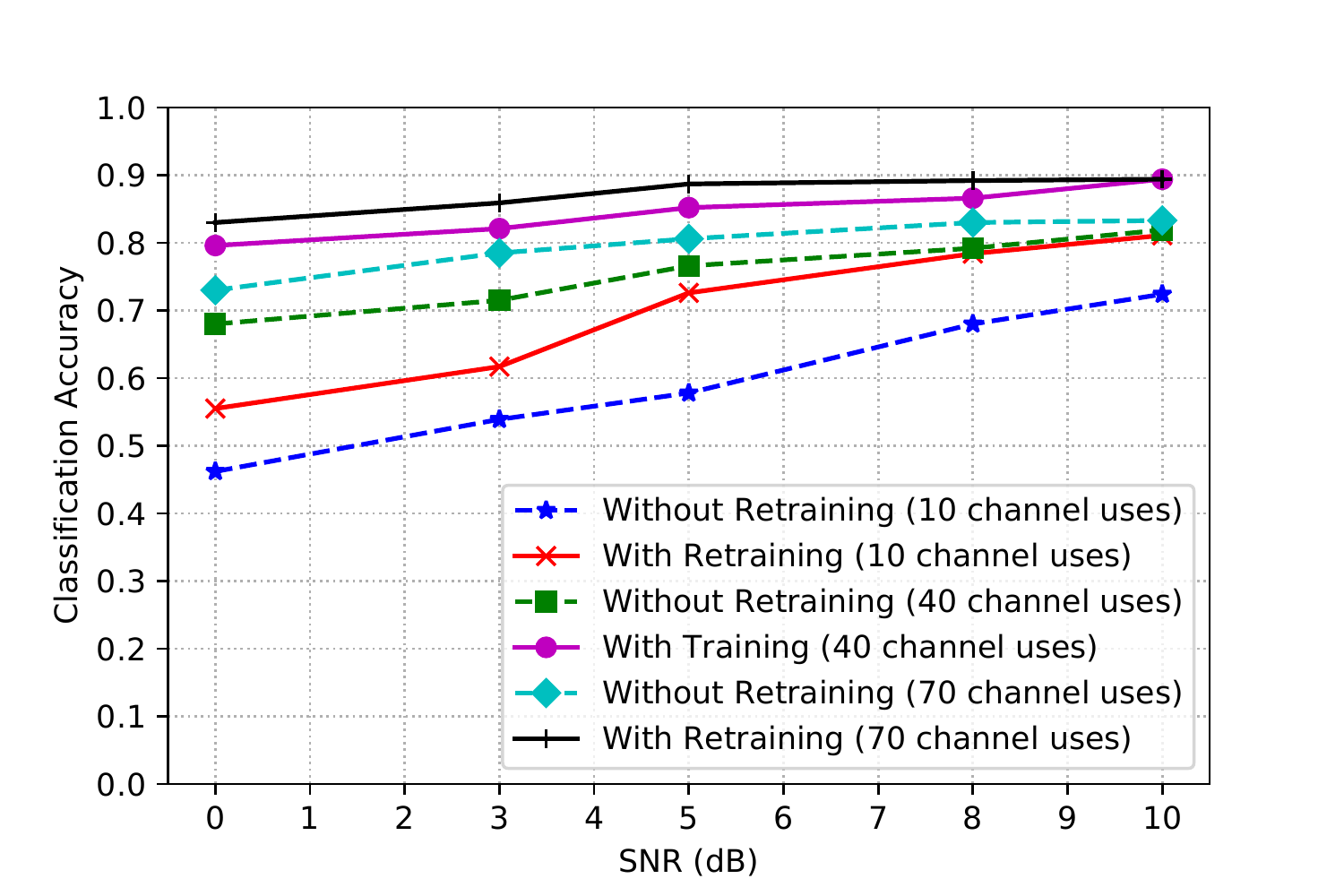}
\caption{Semantic classification accuracy with and without retraining over time.}
\label{fig:trainingresultscombined}
\end{figure}

The DNN architectures of encoder, decoder and semantic task classifier network are provided in Fig. \ref{fig:NNarchs}. Feedforward neural networks are used in each DNN. The MNIST dataset of handwritten digit images (each of $28\times28$ grayscale pixels) is used as the input data (60K for training and 10K for testing). The wireless transmissions are carried out over an additive white Gaussian noise (AWGN) channel. The accuracy of semantic task classifier for both cases of fixed pretraining and interactive retraining over multiple rounds is shown in Fig. \ref{fig:trainingresultscombined} as a function of the SNR (when the number of channel uses is varied). The interactive retraining process helps improve the fidelity of preserving the semantic information and the semantic task classifier performance improves as the SNR and the number of channel uses increase. In the meantime, the reconstruction loss remains small. For example, when the number of channel uses is set to 40, the reconstruction loss remains limited to $0.026$, $0.021$, $0.019$, $0.017$, and  $0.016$ as we vary the SNR as 0dB, 3dB, 5dB, 8dB, and 10dB, respectively.  Overall, semantic communications can effectively reconstruct the input data at the receiver while preserving the semantics of the data.

\section{Adversarial Attacks on Semantic Communications} \label{sec:Attacks}

In test time, \emph{adversarial (evasion) attacks} seek to manipulate the test data input to the adversary's model (e.g., by adding a small perturbation) such that it cannot make a reliable decision for these samples. The effect of this attack can be measured in terms of the model accuracy for the manipulated test input samples (the lower this accuracy drops, the more effective the attack becomes). The perturbation is selected by minimizing the perturbation power subject to the conditions that (i) an error occurs in the decision of the victim model and (ii) the perturbation power remains upper bounded by a threshold. Since solving this optimization problem is difficult, Fast Gradient Method (FGM) can be applied by linearizing the loss function and using the gradient of the loss function when crafting the perturbation. Fast Gradient Sign Method (FGSM) takes the sign of the gradient to design the perturbation. Other attack methods include Basic Iterative Method (BIM), Projected Gradient Descent (PGD), Momentum Iterative Method, DeepFool, and Carlini Wagner (C\&W).

The adversary can launch \emph{targeted} and \emph{non-targeted attacks}. The targeted attacks seek to cause errors in the DNN outputs only for samples from a specific set of non-target labels (classes) to other target labels. On the other hand, the non-targeted attacks seek to cause errors for samples from all labels. The adversary aims to maximize the loss function of the victim DNN for all samples under the non-targeted attack, whereas the adversary aims to minimize the loss function of the victim DNN with respect to the target label under the targeted attack. 

\begin{figure}[h]
\centering
\includegraphics[width=\columnwidth]{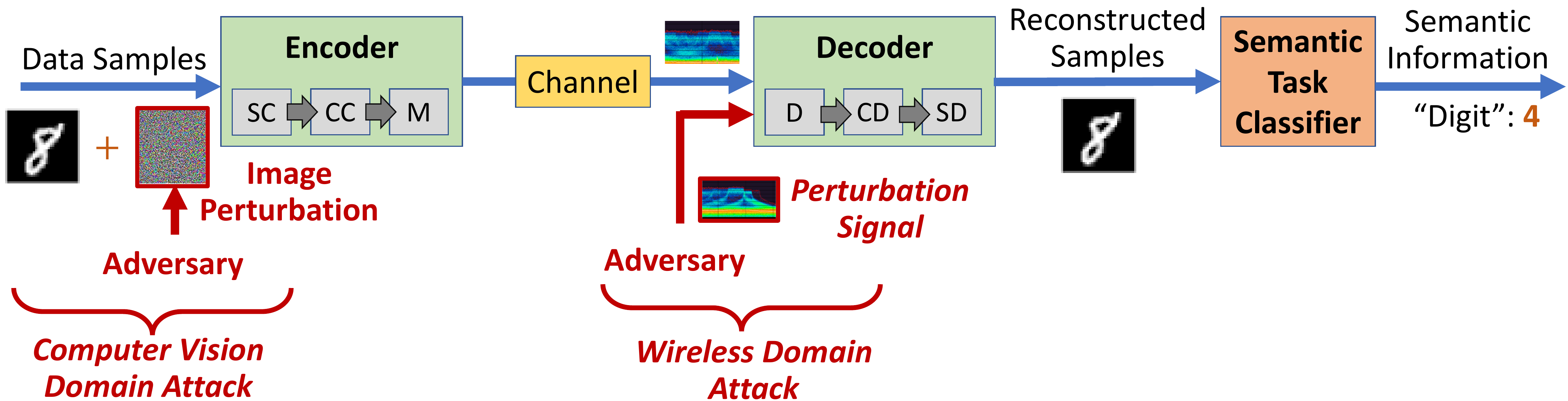}
\caption{Adversarial attacks on semantic communications.}
\label{fig:attacksemantic}
\end{figure}

The adversary can launch adversarial attacks on semantic communications in two different ways. First, the adversary can add a small perturbation to the input sample (namely, the image in our case) and manipulate the semantic meaning of messages (namely, the reconstructed image is classified to a wrong digit label) although the reconstruction loss remains small. Second, the adversary can add a small perturbation to the input of the decoder (namely, the received wireless signal) at the receiver (potentially with an over-the-air transmission). These adversarial attacks in different data domains are illustrated in Fig. \ref{fig:attacksemantic}. 

First, we consider the adversarial attack on the transmitter (encoder) input. The adversarial perturbation is generated by the FGSM by taking the gradient with respect to the concatenation of the autoencoder system and the semantic task classifier. The perturbation is computed as the gradient weighted with the \emph{perturbation strength} and added to the image sample before inputting it to the encoder. This corresponds to a \emph{computer vision attack}. The attack success rate is defined as the average error probability of the semantic task classifier under the non-targeted attack and as the probability that the semantic task classifier classifies a non-target label as a target label under the targeted attack. 

\begin{figure}[h!]
\centering
\includegraphics[width=\columnwidth]{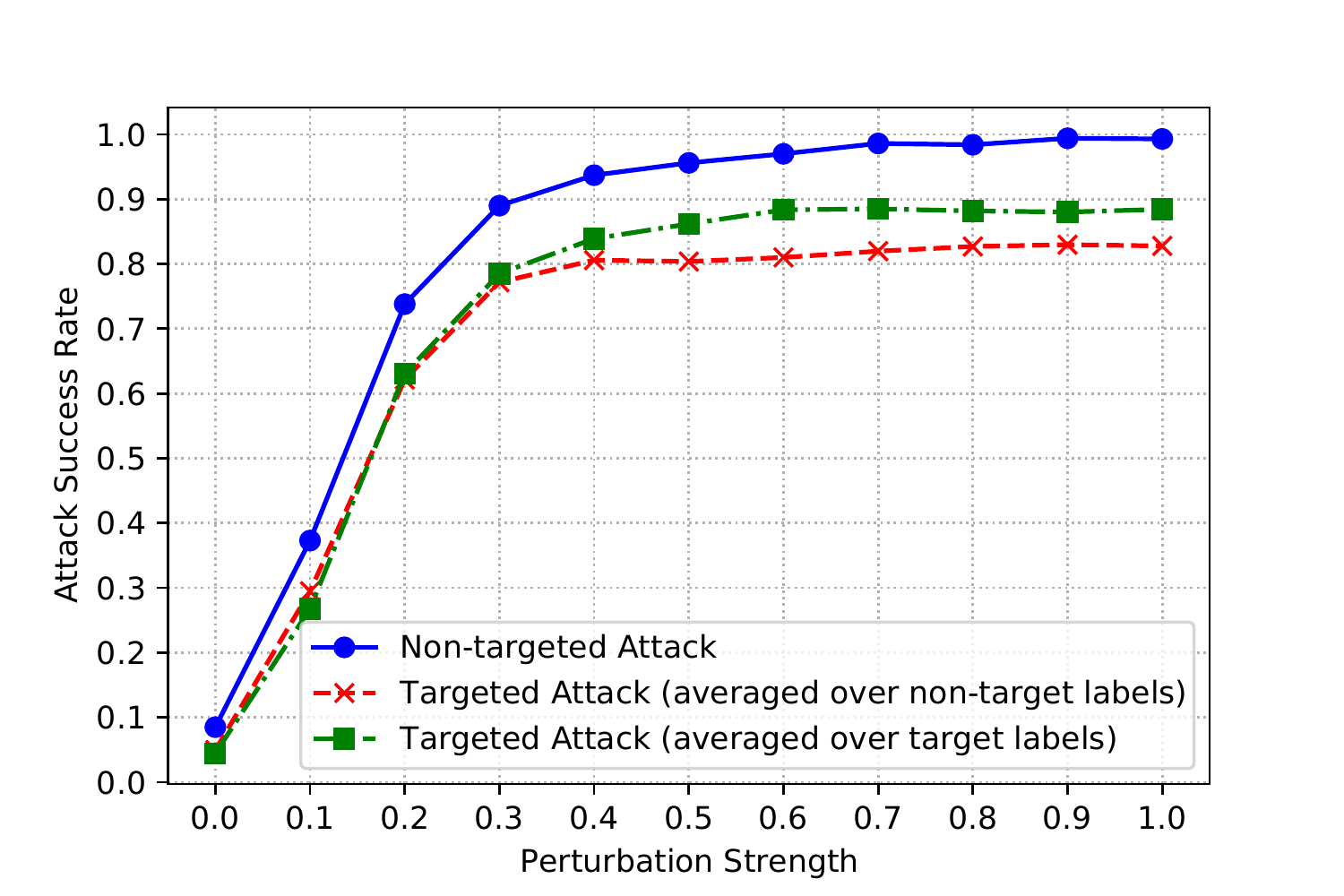}
\caption{Effects of adversarial attacks from the computer vision domain on semantic communications.}
\label{fig:attackresults}
\end{figure}

The success rate of the adversarial attack (at 5dB SNR and with 40 channel uses) is shown in Fig.~\ref{fig:attackresults} as a function of the perturbation strength.  For the targeted attack, we consider two cases; namely, averaged over non-target labels and target labels. We compute the attack success rate for each pair of non-target and target labels. In the former case, we find the best attack success rate over all target labels for a given non-target label, and then average this best attack performance over all target labels. In the latter case, we find the best attack success rate over all non-target labels for a given target label, and then average this best attack performance over all non-target labels. Note that the non-target attack is easier as it needs to flip any label to any other label, so it achieves high attack performance. Targeted attacks are also effective on the average and can likely flip the labels from a specific non-target label to another one or to a specific target-label.

On the other hand, the reconstruction loss caused by each attack is found very close to each other suggesting that adversarial attack poses a more serious threat to the fidelity of semantic information than the information transfer itself. For example, the non-targeted attack with a small perturbation strength such as 0.3 reduces the classifier accuracy to 0.11, although the reconstruction error remains small, namely the MSE is 0.09, suggesting that the adversarial attack can effectively degrade the semantics of the information even when the input data can be reconstructed with a small distortion. For the MNIST data, this means that the images that are reconstructed at the receiver may look similar to the input images but they cannot be reliably classified to its digit labels using the semantic task classifier.          

\begin{figure}[h!]
\centering
\includegraphics[width=\columnwidth]{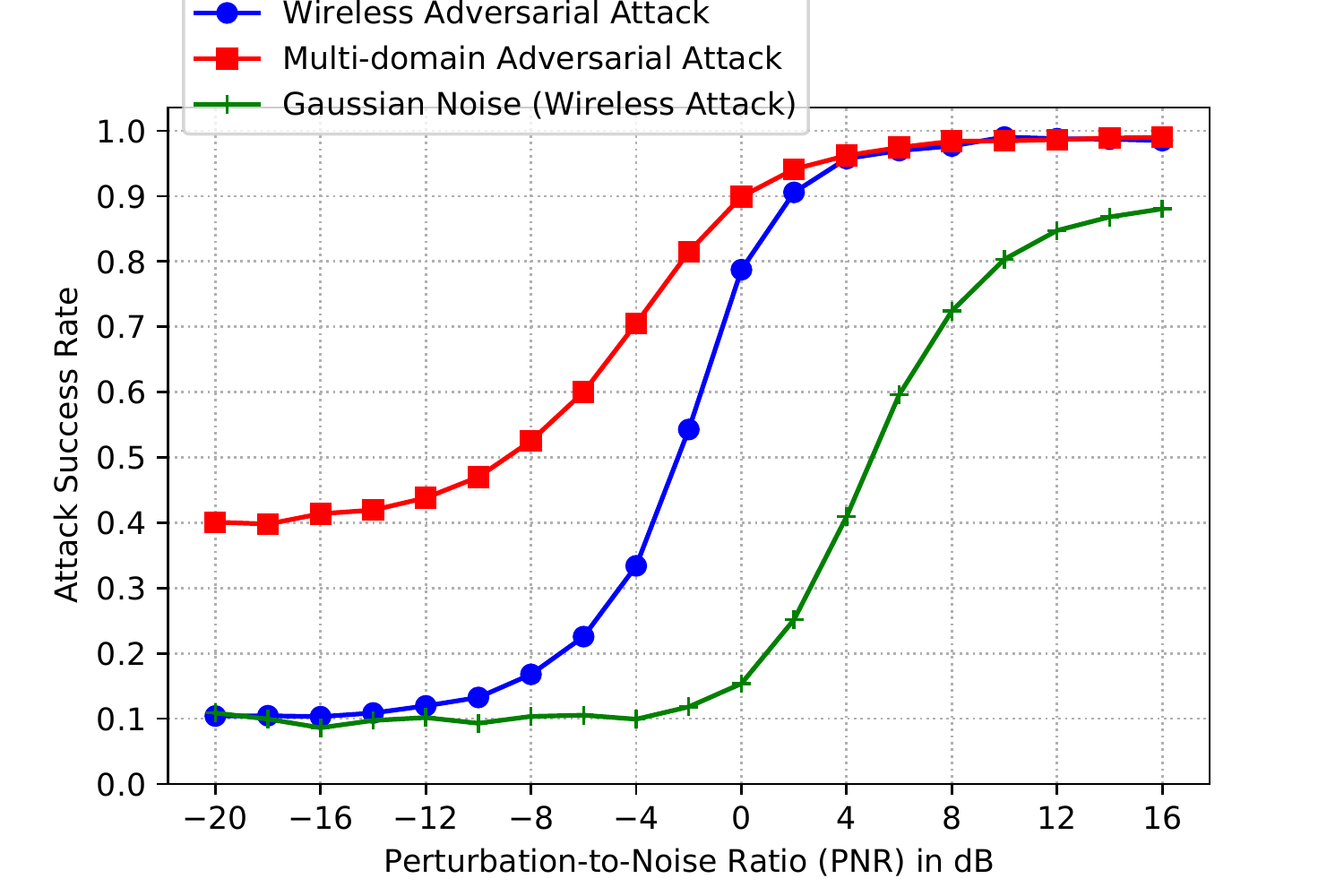}
\caption{Effects of adversarial attacks from the wireless domain domain and multi-domain adversarial attacks on semantic communications).}
\label{fig:combinedattackresults}
\end{figure}

Second, we consider the adversarial attack on the receiver (decoder) input. The adversarial perturbation is generated by FGSM by taking the gradient with respect to the concatenation of the decoder of the autoencoder system and the semantic task classifier. The perturbation is computed as the gradient weighted with the \emph{perturbation-to-noise ratio} (PNR)) and added over the air to the wireless transmission. This corresponds to a \emph{wireless attack}. The success rate of non-targeted attacks on semantic communications (at 5dB SNR and with 40 channel uses) is shown in Fig.~\ref{fig:combinedattackresults}. The performance is evaluated as a function of the PNR and compared with the case when Gaussian noise is used as the perturbation such as in conventional jamming attacks. Results show that the adversarial perturbation added over the transmitted signals is very effective in reducing the classifier accuracy even when the PNR is low. On the other hand, the attack using Gaussian noise as the perturbation is not effective and requires the perturbation power to be much higher than the receiver noise. 

Finally,  Fig.~\ref{fig:combinedattackresults} also shows the performance under the \emph{multi-domain attack}, where a perturbation is added to the receiver input over the air (a wireless attack) and another perturbation (of strength 0.1) is added to the transmitter input image (a computer vision attack). This multi-domain attack combining attacks from the wireless and computer vision domains is the most effective one and quickly reduces the classifier even the adversary uses small perturbations of low strength and PNR. From Fig.~\ref{fig:combinedattackresults}, we observe that the addition of perturbations to the input images does not lose its effect over the wireless transmission when another perturbation is added to the wireless signal. On the contrary, the introduction of the computer vision attack amplifies the effect of the overall adversarial attack and thus further reduces the PNR needed by the wireless adversarial attack to reduce the semantic accuracy below a target threshold.

In this paper, we draw the attention to the security vulnerabilities of semantic communications to adversarial attacks in test time. In adversarial machine learning, it is also possible to attack the DNNs in training time. \emph{Poisoning (causative) attacks} seek to manipulate the training process of the DNNs by manipulating their training datasets in terms of features and labels. These attacks can poison the training datasets for both computer vision and wireless attacks, and their combinations, and the DNNs trained with poisoned data can make errors in both signal reconstruction and recovery of semantic information. Test-time and training-time attacks can be also combined in \emph{backdoor (Trojan) attacks}. For that purpose, the adversary can add triggers to training data (e.g., stickers to images or phase shifts to wireless signals) and modify the corresponding labels. When the DNNs are trained with these triggers, the adversary can activate the triggers by adding them to data samples in test time and cause the DNNs to make errors only for these samples, where the DNNs keep making correct decisions for the unpoisoned test samples. With the open-source development paradigm of O-RAN for next-generation communication systems, these attacks pose a serious threat to the deep learning-driven communication systems such as we discussed for semantic communications. Therefore, efforts are needed to characterize this emerging attack surface and develop defense mechanisms to detect and mitigate these stealth attacks.         

\section{Conclusion} \label{sec:Conclusion}
We formulate an autoencoder-based semantic communications system enabled by deep learning to transfer information from a source to its destination while preserving the semantics of information in addition to reliability objectives. The transmitter and receiver functionalities are represented as the encoder-decoder pair of an autoencoder that is trained with a custom loss function that combines the reconstruction loss with a semantic loss that is captured by the loss of a subsequent semantic task classifier. By accounting for channel effects, the DNNs for the autoencoder and the semantic task classifier are interactively trained to enhance the semantic communications performance. The use of the DNNs makes semantic communications vulnerable to adversarial attacks that seek to manipulate the DNN inputs. We show that these adversarial attacks can be launched in different domains such as a computer vision attack that injects a perturbation to the input image at the transmitter and a wireless attack that transmits a perturbation signal that is received by the decoder at the receiver as superimposed with the transmitted signal. We show that these attacks are very effective individually and even more when combined to reduce the semantic communications performance. In particular, we show that the multi-domain adversarial manipulations can not only increase the reconstruction loss but more importantly lead to a major semantic loss such that the attempt to recover information cannot preserve the semantics.   

\bibliographystyle{IEEEtran}
\bibliography{references}

\end{document}